\documentclass[aps,prl,showpacs,twocolumn,reprint]{revtex4-1}
\usepackage{amsmath,amssymb}
\usepackage{graphicx}
\usepackage[urlcolor=black,colorlinks=true]{hyperref}
\usepackage{multirow}
\begin{document}

\title{Distinction of Nuclear Spin States with the Scanning Tunneling Microscope}
\author{Fabian Donat Natterer}
\author{Fran\c{c}ois Patthey}
\author{Harald Brune}
\affiliation{Institute of Condensed Matter Physics (ICMP), Ecole Polytechnique F{\'e}d{\'e}rale de Lausanne (EPFL), Station 3, CH-1015 Lausanne}
\begin{abstract}
We demonstrate rotational excitation spectroscopy with the scanning tunneling microscope for physisorbed H$_2$ and its isotopes HD and D$_2$. The observed excitation energies are very close to the gas phase values and show the expected scaling with moment of inertia. Since these energies are characteristic for the molecular nuclear spin states we are able to identify the para and ortho species of hydrogen and deuterium, respectively. We thereby demonstrate nuclear spin sensitivity with unprecedented spatial resolution.
\end{abstract}
\pacs{67.63.Cd, 67.80.ff, 67.80.F-, 33.20.Sn, 21.10.Hw, 68.43.-h, 68.37.Ef}
\maketitle

Inelastic electron tunneling spectroscopy (IETS) probes the energies of atomic and molecular excitations in a tunnel junction. When the electron energy reaches the excitation threshold, a new conductance channel opens, leading to a step in the differential conductance $(dI/dV)$. IETS measurements were first carried out in planar tunnel junctions probing vibrations~\cite{lam68} and magnetic excitations~\cite{wol89} of large ensembles of molecules or atoms. A major breakthrough was achieved when performing IETS with the scanning tunneling microscope (STM). This has first been demonstrated for molecular vibrations~\cite{sti98b}, identifying the molecules and their isotopes~\cite{lau99}. A few years later, this was followed by spin-excitations, revealing the Land\'e $g$-factor, effective spin moment, and magnetic anisotropy energy~\cite{hei04,hir07}. In either case, this information is retrieved for individual atoms and molecules of well known adsorption site, coordination-number, and -chemistry. These studies have significantly improved our understanding of surface chemistry and magnetism.

The only process that could so far not be characterized by IETS, neither in planar junctions nor in STM, are true molecular rotations. Albeit, their excitation energies contain manifold information, {\it e.g.}, on chemical identity, bond lengths, rotational degrees of freedom, and molecular conformations. Notably, for homonuclear diatomics, the allowed rotational transitions depend on the nuclear spin state.

Here we demonstrate rotational excitation spectroscopy (RES) with the STM for physisorbed hydrogen, deuterium, and deuterium-hydride. We observe sharp conductance steps in $dI/dV$ at the energies corresponding to the allowed rotational transitions of the respective molecules in the gas phase. The ortho and para nuclear spin isomers of hydrogen and deuterium entail different rotational ground states~\cite{sil80}. We identify their distinct excitation energies and thus demonstrate nuclear spin sensitivity on ensembles containing by many orders of magnitudes less molecules than probed in neutron diffraction~\cite{nie77,fra88}, nuclear magnetic resonance~\cite{kub85,kim97}, and high-resolution electron energy loss spectroscopy (HREELS)~\cite{and82,avo82,pal87}. The mechanism at the origin of STM-RES is proposed to involve a resonant molecular ensemble state. In order to prevent its screening by the metal substrate, we introduced a monolayer of hexagonal boron nitride ($h$-BN) or graphene~\cite{nat13b}.

\begin{figure}
\begin{center}
\includegraphics[width = 8.6 cm]{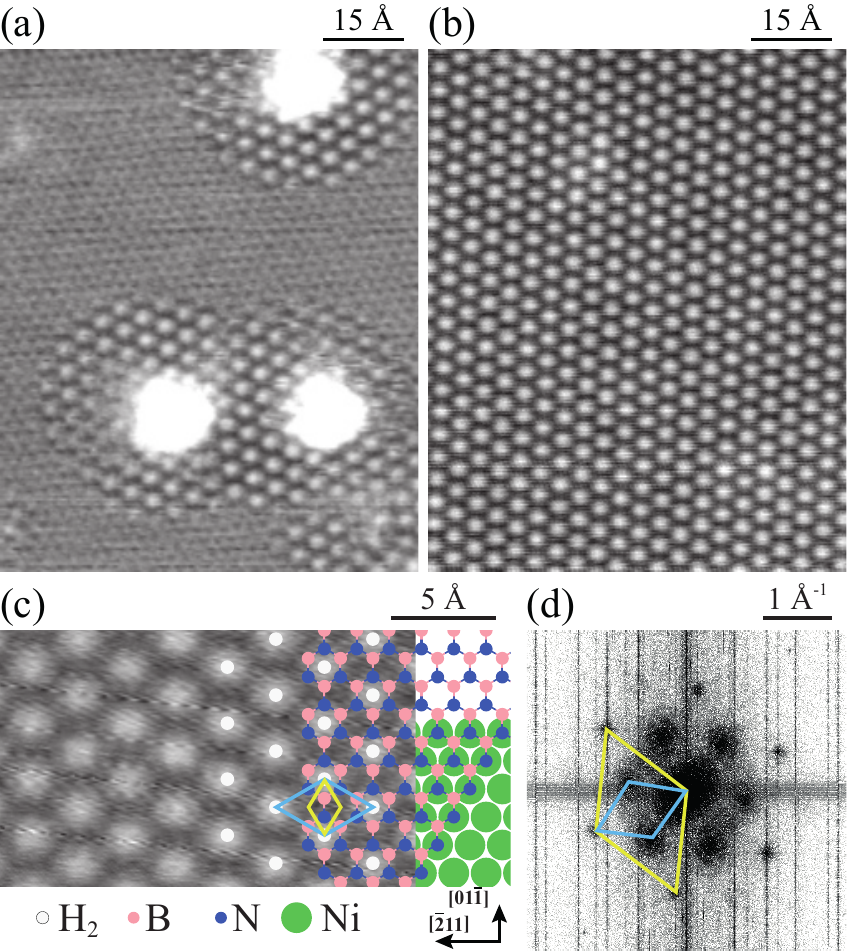}
\end{center}
\vspace{-0.5 cm}
\caption{STM images of H$_2$ superstructure on $h$-BN/Ni(111)--$(1 \times 1)$. (a) Atomically resolved $h$-BN and circular areas of hydrogen superstructure centered around Ti adatoms (exposure 1~Langmuir H$_2$, 1~L~$= 1.33 \times 10^{-6}$~mbar~s, $V_{\rm t} = -10$~mV, $I_{\rm t} = 20$~pA). (b) Full $(\sqrt3 \times \sqrt3) \rm{R} 30^{\circ}$ H$_2$ monolayer ($100$~L H$_2$, $V_{\rm t} = -20$~mV, $I_{\rm t} = 20$~pA). (c) Structure model superimposed on STM image of the H$_2$ superstructure. (d) Fourier transform of (a). The lozenges indicate the $(1 \times 1)$ and $(\sqrt3 \times \sqrt3) {\rm R} 30^{\circ}$ unit cells.}
\label{topo}
\end{figure}

We focus here on molecules that were physisorbed on $h$-BN/Ni(111)--$(1 \times 1)$ grown by chemical vapor deposition using borazine precursors~\cite{nag95}. The H$_2$, D$_2$, or HD molecules were subsequently dosed onto the  surface at 10~K and the STM measurements were performed at 4.7~K. The $dI/dV$ spectra were measured with a Lock-In amplifier using a bias modulation of 2~mV peak-to-peak at 397~Hz.

At low coverages, physisorbed hydrogen forms a two dimensional gas~\cite{nie77} that is transparent to the STM allowing the imaging of the underlying $h$-BN with atomic resolution as shown in Fig.~\ref{topo}. The honeycomb lattice appears as hexagonally close-packed depressions. We adsorbed individual Ti atoms~\cite{nat13a} in order to condense part of the H$_2$ gas in circular islands centered around the adatoms. The H$_2$ molecules are imaged as protrusions. They are in registry with the $h$-BN depressions, however, at $\sqrt 3$ times their distance and rotated by 30${^\circ}$. Upon saturation coverage, H$_2$ forms a perfectly ordered monolayer of this ($\sqrt{3} \times \sqrt{3})$R$30{^\circ}$ superstructure, see Fig.~\ref{topo}~(b). Many weakly physisorbed adsorbates adopt this structure, notably hydrogen on graphite~\cite{nie77,seg82,kub85} and on boron nitride~\cite{kim97}.

\begin{figure*}
\begin{center}
\includegraphics[width = 16.05 cm]{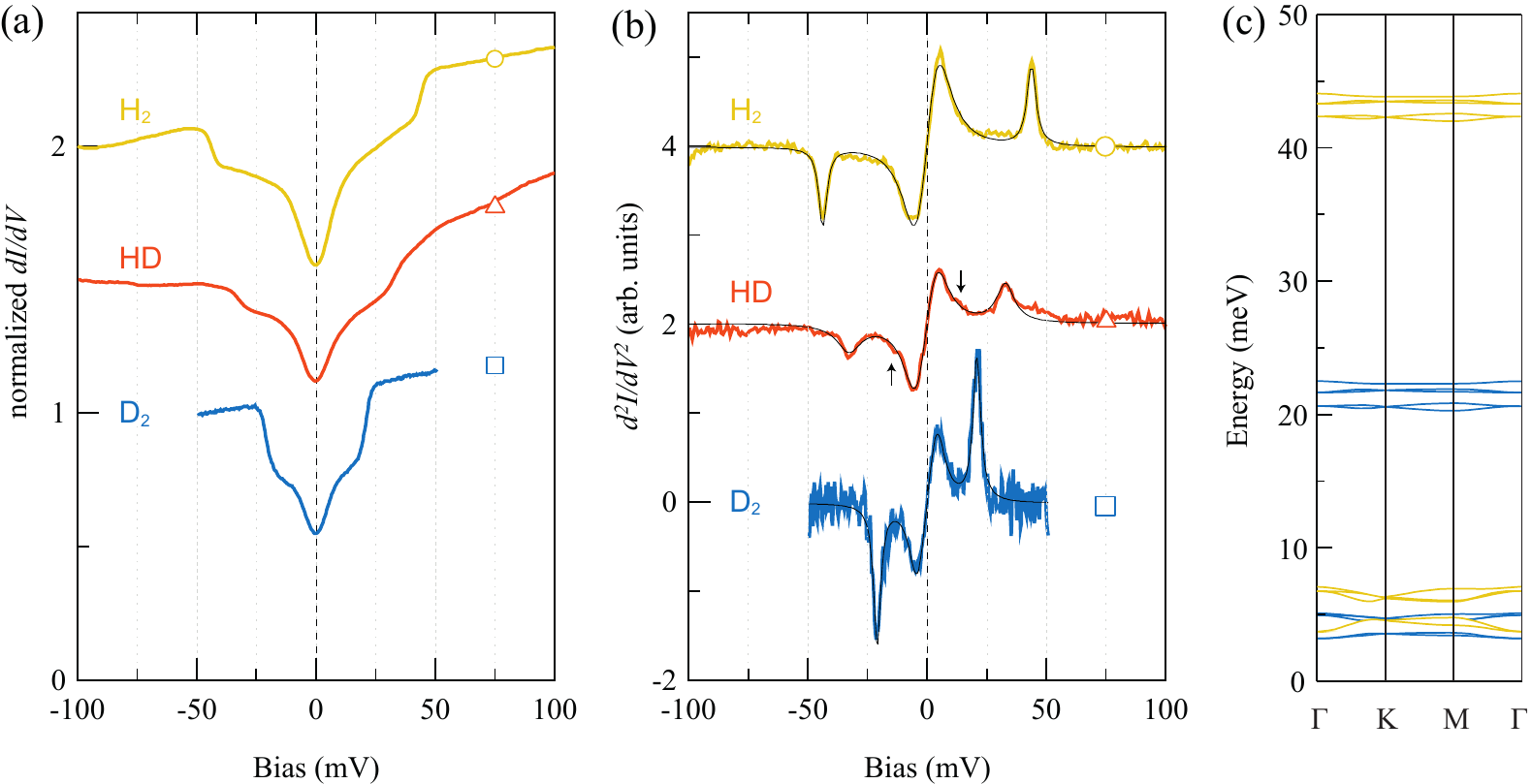}
\end{center}
\vspace{-0.5 cm}
\caption{Rotational excitation spectroscopy on molecular monolayers. (a) $dI/dV$ of H$_2$ $(\bigcirc)$, HD $(\triangle)$, and D$_2$ $(\square)$. The conductance steps at $43.7$, $32.8$, and $20.9$~meV represent rotational $J = 0\rightarrow 2$ transitions. The values for H$_2$ and D$_2$ are characteristic for the para-H$_2$ and ortho-D$_2$ configuration, respectively. The low-energy steps close to the Fermi energy are attributed to phonon gaps of the molecular layer. Spectra are averages of 615 ($\bigcirc$), 1100 ($\triangle$), and 75 ($\square$) $dI/dV$-curves. The spectra were vertically offset by 0.5~nA/V for clarity. (b) Numerical derivative of (a), $d^2I/dV^2$. The black full lines show fits with a multi peak Lorentzian function~\cite{fit}. The shoulder at $(13 \pm 1)$~meV (arrows) for HD is attributed to a $J = 0 \rightarrow 1$ transition. (c) Roton and phonon bands calculated for $p$-H$_2$ and $o$-D$_2$ on graphite~\cite{jan91}.}
\label{didv}
\end{figure*}

The $dI/dV$ spectra on the full monolayers of H$_2$, D$_2$, and HD are shown in Fig.~\ref{didv}~(a). Each curve reveals  two pairs of conductance steps. Their threshold energies are symmetric around zero bias, as expected for IETS. The numerical derivatives $d^2I/dV^2$ in Fig.~\ref{didv}~(b) are used to determine their values. We focus first on the high energy excitations that are located at $(43.75 \pm 0.07)$, $(32.8 \pm 0.4)$, and $(20.89 \pm 0.07)$~meV, for H$_2$, HD, and D$_2$, respectively. Their ratios unambiguously identify them as rotational transitions since the energy of a rotational quantum state $J$ of a linear molecule is with $E_J^{\rm rot} = J(J+1)\hbar ^2/2 I$ inversely proportional to its moment of inertia $I$. In addition, the absolute excitation energies of all three molecules match the reported gas-phase values~(cp. Table~\ref{table1}). Most importantly, the spectra identify H$_2$ in its para and D$_2$ in its ortho nuclear spin configuration.

\begin{table}[h]
\caption{Comparison of gas phase and surface adsorbed rotational excitation energies $\Delta E^{\rm rot}(J\rightarrow J')$ for the three molecules in their vibrational ground state. $S$ labels the molecular spin and $J$ the rotational ground state quantum number. Note that even--odd transitions are forbidden for H$_2$ and D$_2$, while they are allowed for HD. The gas phase $\Delta E^{\rm rot}$ values are taken from Ref.~\cite{sil80}. The error bars for STM-RES indicate the standard deviation.}
\label{table1}
\vspace{0.5cm}
\centering
\begin{tabular}{|ccccccc|}
\hline
\multicolumn{2}{|c}{Molecule} &$S$ & $J$   & \multicolumn{2}{c}{$E^{\rm rot}$ (meV)}  & STM-RES (meV)\\
\multicolumn{4}{|c}{ } & $\Delta J=1$  & $\Delta J=2$ & \\ \hline \hline
\multirow{2}{*}{H$_2$}& para& 0 & 0     & $-$          & 43.9 &  $43.75\pm0.07$\\
& ortho & 1 & 1     & $-$          & 72.8 &  not observed\\ \hline \hline
HD  &      &  & 0     & 11.1        & 32.9 & $13\pm1$ and $32.8\pm0.4$\\ \hline \hline
\multirow{2}{*}{D$_2$}& ortho &0/2 & 0     & $-$          & 22.2 & $20.89\pm0.07$ \\
&para &1 & 1     & $-$          & 36.9 & not observed\\ \hline

\end{tabular}
\end{table}

The distinct rotational excitation energies of the disparate nuclear spin states are caused by symmetry constraints of the total molecular wavefunction~\cite{sil80}. It is a product of the nuclear, rotational, electronic, and vibrational wavefunctions. Hydrogen nucleons are fermions, therefore this product must be antisymmetric with respect to proton permutation. For hydrogen, the vibrational and the electronic ($^1\Sigma^+_g$) ground states are symmetric. Consequently, the antisymmetric nuclear singlet state ($S = 0$, para) requires a symmetric rotational wavefunction (even $J$), whereas the symmetric nuclear triplet state ($S = 1$, ortho) implies an antisymmetric rotational wavefunction (odd $J$). The nucleons of deuterium are bosons requiring a symmetric molecular wavefunction. Hence, the symmetric nuclear spin state is associated with a symmetric rotational state and  the antisymmetric nuclear spin configuration with an antisymmetric rotational state. Transitions between the nuclear spin isomers are forbidden for free molecules, but are catalyzed by paramagnetic impurities or inhomogeneous magnetic and electric fields~\cite{ili92,sug11}.

For the case of HD, the nucleons are distinguishable and the above symmetry constraints do not apply. The observed threshold energy for HD is in agreement with a $J = 0 \rightarrow 2$ excitation~(Fig.~\ref{didv}). For this molecule, also $\Delta J = 1$ transitions are allowed, and the spectrum indeed shows a little shoulder at $(13 \pm 1)$~meV close to the reported $J = 0 \rightarrow 1$ transition energy (Table~\ref{table1}). Note that the RES steps are with 11 -- 37~\%~\cite{fit} significantly higher than the ones of vibrational-excitations for adsorbates on metals~\cite{sti98b}, and they are comparable with the spin-excitation step heights observed for magnetic atoms on a decoupling monolayer~\cite{hei04,hir07}.

We attribute the low-energy conductance steps in Fig.~\ref{didv} to the excitation of phonons in the molecular layers. As can be seen from Fig.~\ref{didv}~(c), the substrate potential creates a phonon gap at the Brillouin zone center reaching from zero to the energy where the weakly dispersing bands are located~\cite{fra88,nov88,lau89,jan91}. This creates a narrow energy interval in which phonons can be excited and thus meets the necessary condition for the observation of a distinct threshold energy in IETS. The excitations are with $(5.5 \pm 0.5)$, $(5.1 \pm 0.5)$, and $(4.4 \pm 0.5)$~meV, for H$_2$, HD, and D$_2$, remarkably close to the centers of the measured phonon bands~\cite{fra88,lau89}. Notably, the H$_2$/D$_2$ energy ratio of 1.3 matches the one of the phonon gaps. The deviation from $\sqrt{m_{\rm D_2}/m_{\rm H_2}} = \sqrt{2}$ is caused by the anharmonicity of the intermolecular and of the adsorption potential~\cite{fra88,nov88,jan91,mat80}.

\begin{figure}
\begin{center}
\includegraphics[width = 8.6 cm]{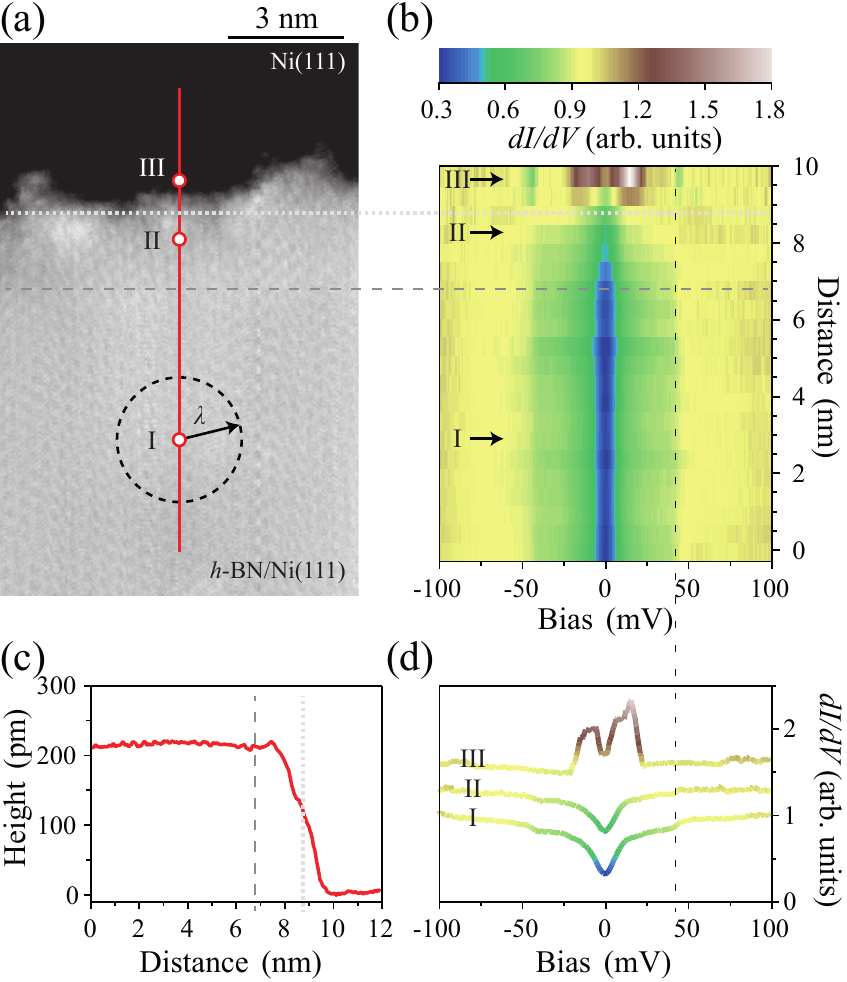}
\end{center}
\vspace{-0.5 cm}
\caption{The lateral extent $\lambda$ of the rotational excitation. (a) STM image showing a step between the $h$-BN layer and the underlying Ni substrate, both are H$_2$ covered. (b) $dI/dV$ taken along the full red line perpendicular to the step. The rotational excitation is attenuated at $\lambda = (1.7 \pm 0.5)$~nm from the step, see gray dashed line. (c) Line-profile across the step in (a). (d) $dI/dV$ spectra at the indicated positions I, II, and III. Spectra represent averages of 16 $dI/dV$-curves.}
\label{range}
\end{figure}

We determined in Fig.~\ref{range} the lateral range $\lambda$ of excited molecules probed with STM-RES by recording $dI/dV$ spectra on samples with partial $h$-BN coverage. This gives access to physisorbed H$_2$, once in the $\sqrt{3}$ phase on $h$-BN, for which we observe RES features, and once directly adsorbed on Ni(111), where rotational excitation conductance steps are absent, in agreement with former low-$T$ STM studies~\cite{gup05,tem08,sic08b,wei10}. The $dI/dV$ spectra recorded across the transition between these two surface terminations reveal the attenuation of the rotational excitation when approaching the $h$-BN border to a distance of $\lambda = (1.7 \pm 0.5)$~nm. Within this radius there are $(60 \pm 30)$ hydrogen molecules. This represents the ensemble size probed for the present system by STM-RES.

The mechanism underlying STM-RES has to be different from the one of HREELS, since the latter detected rotational excitations for physisorbed hydrogen on metal surfaces~\cite{and82,avo82}, while in STM there were no signs of such excitations for the same systems~\cite{gup05,tem08,sic08b,wei10}. Instead, all observed spectroscopic features were reminiscent of two state switching~\cite{gup05}. In EELS, a negative ion resonance is populated by the primary electrons and subsequently decays into several inelastic channels, one of them being the molecular rotation~\cite{sch73,pal92}. For hydrogen this resonance is at an energy accessible to the typical 5~eV incident electrons~\cite{sch73,and82,avo82,tei02}, but evidently not to IETS operating at electron energies near the excitation threshold of the molecular rotations. However, the existence of a rotational resonance for ultra-low electron energies was demonstrated by molecular density dependent electron drift velocity measurements~\cite{fro68,cro71}. It has been attributed to a collective resonance state originating from the electrostatic, polarization, and quadrupole interactions between neighboring hydrogen molecules~\cite{gar77}. This state requires molecular densities of comparable order of magnitude than in the present work ($6.2 \times 10^{14}$~cm$^{-2}$). Similar densities were also present in former STM studies~\cite{gup05,tem08,sic08b,wei10} and the absence of RES features, that we also note in Fig.~\ref{range} for H$_2$/Ni(111), must be due to screening of the intermolecular interactions by the underlying substrate. Therefore a decoupling $h$-BN layer enables the spectroscopy of molecular rotations with STM-IETS and the collective nature of the resonant state explains the finite lateral range of excited molecules. According to this mechanism, STM-RES is an intrinsically multi-molecule method, requiring for the case of H$_2$ at least 60 interacting molecules. Our collective rotational excitations can be interpreted in terms of the calculated roton bands~\cite{jan91} shown in Fig.~\ref{didv}~(c). We note that 60 molecules is an unprecedented small number in terms of the demonstrated access to the nuclear spin state.

With para-H$_2$ and ortho-D$_2$ we observe for each molecule only the nuclear spin isomer with lowest energy rotational ground state, $J = 0$. However, for either molecule both nuclear spin isomers are present in the gas phase, with ortho/para ratios of $3/1$ for H$_2$ and $2/1$ for D$_2$ at room temperature. Therefore a fast conversion to the lowest energy nuclear spin configuration must take place on the surface. In line, former EELS studies either observed only the spin isomer with even $J$~\cite{pal87}, as in our case, or they reported the ortho to para-H$_2$ conversion after a few minutes~\cite{avo82}. This conversion has been attributed to short range magnetic interactions with the surface~\cite{ili92}. We never observed ortho-H$_2$ and para-D$_2$, neither on $h$-BN nor on graphene, both grown on non-magnetic substrates~\cite{nat13b}. Therefore the conversion to the lowest energy configuration must be driven by magnetic impurities. The electric fields of the tunnel junction are by at least one order of magnitude smaller than required~\cite{sug11}.

There is pioneering STM work related to molecular rotations. STM was used to induce and monitor the rotation of O$_2$/Pt(111)~\cite{sti98a}, the coupling of vibrational and rotational degrees of freedom was demonstrated for acetylene/Cu(100)~\cite{sti98c}, and hindered rotations were reported for CO on two low-index Cu surfaces~\cite{lau99}. However, here we reveal for the first time the molecular rotational eigenvalues and thereby complete the meanwhile well established and widely used STM-IETS vibrational and spin-excitation spectroscopy.

We demonstrated for H$_2$ and D$_2$ nuclear spin sensitivity and proposed a mechanism involving a collective low-energy resonant state that emerges from molecular interactions. STM-RES gives access to the eigenvalues of any molecular rotor; particularly well suited are molecules with large rotational constants $\hbar ^2/2 I$. For homonuclear diatomics, such as N$_2$ and O$_2$, the nuclear spin states can now be inspected with unprecedented spatial resolution, as well as the intriguing ordering phenomena of ortho--para mixtures~\cite{kub85,kim97}. Furthermore, the coupling of the nuclear spin to the atomic environment and potentially also nuclear processes become accessible on a local scale. We finally note that even single molecule STM-RES might be feasible for those molecules with very low and broad negative ion resonance energy, thereby creating the necessary overlap with the rotational excitation threshold.

Funding from the Swiss National Science Foundation is greatly appreciated.
\bibliography{ms_10}

\end{document}